\documentclass[aps,pra, onecolumn,14 pts,a4paper,showpacs]{revtex4}

\usepackage{amssymb,amsmath,amsfonts,graphicx}
\usepackage{bm}

\begin{document}

\title{Wave-breaking and generic singularities of nonlinear hyperbolic equations}

\author{ Yves Pomeau }
\affiliation{Los Alamos National Lab, CNLS, Los Alamos, NM 87545
USA.}
\author{Martine Le Berre}
\affiliation{Laboratoire de Photophysique Mol\'eculaire, Bat.210,
91405 Orsay, France.}

\author{ Philippe Guyenne}
\affiliation{Department of Mathematical Sciences, University of
Delaware, Newark DE 19716-2553, USA.}

\author{ Stephan Grilli}
\affiliation{Department of Ocean Engineering, University of Rhode
Island, Narragansett, RI 02882, USA.}

\date{\today }
\begin{abstract}
\textbf{Abstract}
 Wave-breaking is studied analytically first and the results are compared with accurate
 numerical simulations of 3D wave-breaking.
We focus on the time dependence of various quantities becoming
singular at the onset of breaking. The power laws derived from
general arguments and the singular behavior of solutions of
nonlinear hyperbolic differential equations are in excellent
agreement with the numerical results. This shows the power of the
analysis by methods using generic concepts of nonlinear science.
\end{abstract}

 \maketitle

\section{Introduction}
Among the myriad of natural phenomena dominated by nonlinearity,
wave-breaking is quite special. It is ubiquitous along most oceanic
shorelines, and although progress has recently been made in its
understanding, its full  explanation remains somewhat shrouded in
mystery. This makes wave-breaking an outstanding topic in the study
of nonlinear phenomena and, we believe, one that is particularly
well suited  to this ``Open problems section of Nonlinearity''.
Wave-breaking also plays  a significant role in the larger scale
dynamics of the Oceans \cite{dias}, particularly with respect to
their interaction with the atmosphere. For instance,
 the swell provides nucleation seeds for the rain and wind
 momentum exchange
 with water,  mostly through nonlinear processes. Among such
 processes, wave-breaking could also be
  significant, if not dominant, to
 limit the growth of the instability responsible of the formation of waves.

The linear theory of waves has a long and interesting history, going
back to the Principia. In book 2
 Newton  shows that the velocity of water waves in the infinite depth limit is
 proportional to $\sqrt{g \lambda}$, with
  $g$ the acceleration of gravity and $\lambda$ the wavelength.  Years later, Bernoulli
   derived the shallow water wave velocity, applicable to waves with a wavelength much
  larger (typically more than 20 times) than the water depth. The interpolation between  these two  limits
   is due to Lagrange. Later, in a very
   important paper, Stokes defined the range of applicability of the linear theory in the shallow water
    limit. Thereafter Riemann's method of solution of quasilinear equations
    showed
    the occurrence of a finite time singularity for simple
    waves in the shallow water case. Among the open problems posed by the occurrence of such
    a singularity, we shortly discuss below
     the two existing methods proposed to regularize the wave profile to avoid the
     occurrence of an overhang. These methods introduce the next order space derivative in the shallow
     water equations (i.e., so-called dispersive effects) and thus
     preclude wave-breaking from occurring.

The relationship between wave-breaking and the nonlinear structure
of the governing equations is reconsidered below, where we show that
strong nonlinearities dominate the wave dynamics, both in the
transverse and longitudinal directions.
 Starting from generic
equations, whose solution can develop discontinuities (inviscid
Burgers, or pressure-less fluid equations), we first show in a
one-dimensional space (1D), that a very simple (once known as
\cite{Poisson}) solution $u(x,t)$ of this partial
 differential equation does yield a wave-breaking (i.e., overturning)
 process. Defining $u$ as the free surface elevation, we study
 its dynamical behavior and, more  specifically, the
  growth of the overhang domain with time.

 Extending the pressure-less fluid equation to  a two-dimensional  space  (2D), we obtain a
 generic  equation for $u(x,y,t)$ showing how the ``overhanging
 domain'' of the
 overturning wave extends  in both spatial directions, parallel and normal to the wave.
 In that case, the boundary of the overhanging region, referred
 to as the ``apparent contour'' (i.e., the set of points
  where the tangent to the water surface is vertical), becomes a skewed
  curve, when drawn on the water
  surface. We find that this curve spreads as $t^{3/2}$ in the direction of
  wave propagation and  as
   $t^{1/2}$ along the wave crest ($t =0$ being the instant where
   wave overturning is initiated, yielding an overhanging region).
   These results will be shown to be in good
   agreement with those of numerical simulations of a 3D overturning
   wave over a sloping underwater ridge, based on the
   solution of  Euler
    equations with free surface boundary conditions \cite {GG}. Moreover the lateral spreading as $t^{1/2}$
   was recently found experimentally\cite{marseille} for waves generated by a wave-maker, and
    progressing on an horizontal bottom.

 These time evolution laws are established first for a pressure-less
fluid and then are extended in section
\ref{sec:wavebreakinginshallow} to the shallow water case. The
scaling laws for wave-breaking, in the  directions perpendicular and
parallel to the crest, will be derived  from solutions of
fluid-mechanical equations. It is true that this  wave-breaking
process could be seen as an example of a metamorphosis  of a surface
studied in catastrophe theory, named a 'sickle' or 'lips'. In
particular, Arnold \cite{arnold}, referring to physical situations
completely different  of ours, notices that the width of a 'sickle'
should grow generically like $(t - t_0)^{1/2}$, $t_0$ being the time
of the first singularity. This law of growth is the same as  the one
we find for the lateral spreading of the overturning domain. However
catastrophe theory is very general and cannot be considered as
providing any kind of mathematical method for solving the equations
of fluid mechanics. Therefore we did not feel useful to use this
theory explicitly in the present work and we relied instead on
explicit solutions of the fluid equations. Moreover, to the best of
our knowledge, catastrophe theory has never been used in
investigations of the wave-breaking problem.
This being a contribution to the Open problems section of the
Journal, in section 4 we comment on various aspects of
the problem that do wander off the main track. For instance, we
suggest an extension of the boundary integral method (e.g., such as
used in \cite{GG}) to deal with the wave-breaking problem, beyond
the time when the plunging jet impinges on the free surface
underneath, and briefly discuss  some features of what we call the
Hertz-fluid contact problem.

\section{From 1D to 2D wave-breaking}
\label{sec:from1D2D}

In this section, we first introduce  a generic singular solution to
a nonlinear hyperbolic partial differential equation (PDE), in a
one-dimensional space (1D). We then show that such solutions of the
PDE can be represented by a curve that is continuously deforming
and, hence, can be used as a simple geometric model for predicting
the occurrence of an overhanging region, similar to the plunging jet
in an overturning (i.e., breaking) water wave. We finally extend
this equation to a two-dimensional space (2D), which is a more
physically relevant case.

 \subsection{Wave-breaking in one dimension}

Consider the equation
\begin{equation}
\frac{\partial u}{\partial t} + u \frac{\partial u}{\partial x} = 0
\mathrm{.} \label{eq:1}
\end{equation}
It has \cite{Poisson} the implicit solution
\begin{equation}
 u(x,t) = u_0 (x - u t) \mathrm{,}
\label{eq:2}
\end{equation}
where $u_0 (x)$ is the initial condition. Let $f(.)$ be the inverse
function of $u_0 (.)$ (all functions
 assumed smooth). Therefore:
\begin{equation}
 f(u(x,t)) = x - u t \mathrm{.}
\label{eq:3}
\end{equation}

 This solution $u(x,t)$, as given by equation (\ref{eq:3}), becomes multivalued whenever
 $\frac{\partial x}{\partial u} =0$, the derivative being taken at constant $t$.
 From equation (\ref{eq:3}), this yields
 \begin{equation}
 t + \frac{{\mathrm{d}}f}{{\mathrm{d}}u}  = 0
 \mathrm{.}
 \label{eq:3a}
 \end{equation}
  The smallest $t$ when this has a root, for a given $f(u)$, is when this root is stationary with respect
  to $t$, that is when ${\mathrm{d}}t =0$, with $t$ and $u$  related by (\ref{eq:3a}). This yields
   ${\mathrm{d}}t = - {\mathrm{d}}u \frac{{\mathrm{d}}^2 f}{{\mathrm{d}}u^2}=0$, or simply
   $\frac{{\mathrm{d}}^2 f}{{\mathrm{d}}u^2} = 0$. Therefore, at the singularity, $f$ has a vanishing
    second derivative, a standard result.
  Furthermore it can also be assumed  that the Taylor series
  expansion (TSE) of $f$ has no linear term (amounting to take
   $u =0$ as the value of $u$ where $f(.)$ has an inflection point), since a constant $u$ can be absorbed
   into a redefinition of the time origin in equation (\ref{eq:3}).
   Let us then define $t=0$ as the first time when the
solution becomes singular,  to avoid having to write $(t
-t^*)^{3/2}$ etc. instead of $t^{3/2}$, with $t^*$ time of the
singularity.
  Near the singularity (assumed to occur at $t =0$) one can thus
  expand (after convenient scaling)$f(u)$
  like $ f(u) = - u^3 + a u^4 +...$, with $a$ finite constant. Once put into equation (\ref{eq:3})
  this gives to the 4th-order:
\begin{equation}
 a u^4  =  u^3 + x - u t \mathrm{.}
\label{eq:4}
\end{equation}
For short times , Equation (\ref{eq:4}) implies that $u$ scales like
$t^{1/2}$ and $x$ like $t^{3/2}$, which makes all terms on the
right-hand side of Equation (\ref{eq:4}) of the same order,
$t^{3/2}$. The left-hand side term is - with the same scaling law -
of order $t^2$ and therefore negligible compared to the right-hand
side; this applies as well  to higher-order terms in the TSE of
$f(u)$ near the inflection point. Assuming that the beginning of an
overhanging region can be described by the lower-order terms, the
'universal' equation describing 'wave-breaking'  reads:
\begin{equation}
0  =  u^3 + x - u t \mathrm{.} \label{eq:4a}
\end{equation}
Note that, in this derivation, the coefficient of $u^3$ of the TSE
of $f(u)$ near zero is negative and  thus a singularity will appear
for positive times. For negative times, Equation (\ref{eq:4a})  only
has one real root while it has three such roots for positive times.

Therefore the curve $u(x,t)$ is everywhere single-valued  if its
slope  is negative at $x =0$, which happens for $t$ negative. By
contrast, $u(x,t)$ is triple-valued in a range of values of $x$, if
$t$ is positive. This solution can be put in a self-similar form by
eliminating one variable. The scaling laws for the transition from
single to triple-valued results are $ u\sim\sqrt{|t|}$ and $ x \sim
|t|^{3/2}$. Therefore, the 'natural' solution of Equation
(\ref{eq:4}) is of the form $ u(x,t)  = \sqrt{|t|} U(\zeta)$, with
 $\zeta = x |t|^{-3/2}$ and $U$ a numerical function. Nevertheless, this is not
 sufficient, because the
 function $U(.)$ takes different values depending on whether $t$ is negative,
 zero, or positive. This can
 be taken into account formally by introducing a discrete argument in the function $U(.)$, namely
 an index $i$ with three possible values $-1$, $0$ and $1$, such that $ i =-1$ if $t$ is negative,
  $ i =0$ if $ t =0$ and $ i = 1 $ if $t$ is positive. This defines three real numerical functions
  $U_{i} (\zeta)$, such that $U_{-1}(\zeta)$ is the real root of $ 0 =  U_{-1}^3 +\zeta + U_{-1}$,
    and $ U_0 = - \zeta^{1/3}$. The multivalued function  $U_{1}$ is any real root of
   $ 0 =  U_{1}^3 +\zeta - U_{1}$,
   with  three possible values in the range $-\frac{2\sqrt{3}}{3}<\zeta<\frac{2\sqrt{3}}{3}$ and one
   otherwise.

This completely defines  the self-similar solution.

\subsection{Wave-breaking in more than one space dimension}
\label{subsec:wavebreakin2D}

Our approach to the formation of singularities in solutions of
hyperbolic equations can be extended to nonlinear differential
equations with more than one spatial variable. A simple extension of
the nonlinear equation (\ref{eq:1}) provides a model for the
occurrence of
 singularities for time-dependent  functions of two  coordinates of space
 $(x,y)$.
Let $u(x,y,t)$ and $v(x,y,t)$ be the two Cartesian horizontal
components of the velocity field of a fluid, with $(x,y)$  the
horizontal coordinates and $x$  in the direction of propagation.
Momentum equations, without pressure or gravity read:

 \begin{equation}
 \left \{ \begin{array}{l}
\frac{\partial u}{\partial t} + u \frac{\partial u}{\partial x} +  v \frac{\partial u}{\partial y}= \epsilon \nabla^2 u\\
\frac{\partial v}{\partial t} + u \frac{\partial v}{\partial x} +  v
\frac{\partial v}{\partial y}= \epsilon \nabla^2 v \mathrm{,}
\end{array}
\right. \label{eq:6}
\end{equation}

 with $\epsilon$ a viscosity coefficient. Note that mathematical
models with a similar formalism have been proposed \cite{zeldo} to
describe the formation of large scale structures in the Universe,
assuming a potential flow, namely that a function $\Phi(x,y)$ exists
such that $ u = - \frac{\partial \Phi}{\partial x}$ and $ v = -
\frac{\partial \Phi}{\partial y}$.  Such a dependence of the
velocity field on a potential is possible for the present model: if
the flow is potential at $t =0$, equations (\ref{eq:6}) reduce to
the unsteady Bernoulli equation, which ensures that the flow remains
potential for all times. In this potential flow case, equations with
the viscous terms can still be solved thanks to Florin's
\cite{florin} transform (often referred to as Hopf-Cole transform) :
$S = -2 \epsilon \ln|\Phi|$ , which yields a linear diffusion
equation for $S(x,y,t)$.

For $\epsilon =0$ equations(\ref{eq:6}) have the formal solution very
similar to equation (\ref{eq:3}):
\begin{equation}
\left \{ \begin{array}{l}
x - t u(x,y,t) = F(u,v)\\
y - t v(x,y,t) = G(u,v) \mathrm{} \label{eq:FG}
\end{array}
\right.
\end{equation}

where the functions $F,G$ are presented below.
 Let $M$ be the general two-by-two matrix of the linear group, with non-zero determinant.
  Therefore $(\tilde{u}, \tilde{v}) = M (u,v)$ and $(\tilde{x}, \tilde{y}) = M (x,y)$
are solutions of equations of the same form as the left-hand side of
 Equation (\ref{eq:FG}). This holds true with the {\it{same}} matrix
acting on $(u,v)$ and $(x,y)$. The same general linear map will
change in general the functions $F(u,v)$ and $G(u,v)$ in a rather
complicated way. However because the map is linear it will not
change the degree of a polynomial in $u$ and $v$. Based on that
property, we shall be able to show, almost without any calculation,
that singularities yielding discontinuities of derivatives with
respect to $x$ may spread in any direction,  the analysis being made
by assuming first that the
 singularity spreads normal to $x$.

The functions $F(u,v)$ and  $G(u,v)$ are defined by the initial
conditions. For smooth initial data they are such that the mapping
from $(u,v)$ to $(x,y)$, defined by
 $x = F(u,v)$ and  $ y = G(u,v)$ is one to one. Therefore the Jacobian
  $J(t=0) = \frac{\partial G}{\partial v}\frac{\partial F}{\partial u} - \frac{\partial G}
  {\partial u}\frac{\partial F}{\partial v}  $ never vanishes. The solution of equations
  (\ref{eq:FG})
becomes singular whenever the time-dependent Jacobian $J(t)$ of the
mapping of $(u,v)$ onto $(x,y)$ becomes singular. This Jacobian
reads :
\begin{equation}
J(t) = \left(\frac{\partial G}{\partial v} + t \right)
\left(\frac{\partial F}{\partial u} + t \right) - \frac{\partial
G}{\partial u}\frac{\partial F}{\partial v} \mathrm{.} \label{eq:9a}
\end{equation}
Let us assume that the singularity first occurs  at $t =0$ and
$x=y=0$, with $u=v=0$ as well. The latter restriction is not  as
important as one could believe at first : the original equations are
Galilean invariant, so that one can always specify that the velocity
field is $u=v=0$ at some point of time and space.

The condition equivalent to the cancelation in 1D of the second
derivative of $f(u)$ at $ u=0$, here, is the property that, near $t
= u = v = 0$, the first relevant terms in the TSE of the
 Jacobian $J(t)$ read

\begin{equation}
J(t) = a t + b u^2 + c v^2 + 2d u v \mathrm{,} \label{eq:9}
\end{equation}
where $a$, $b$, $c$ and $d$ are constant derived from the TSEs of
$F(u,v)$ and   $G(u,v)$ near zero. The expansion (\ref{eq:9})
implies that there is no term linear with respect to $u$ and $v$.
Otherwise the Jacobian would vanish for $t$ negative and for small
values of $u$ and $v$, which is not the case
 based on our assumptions. The two coefficients of the part of the
Jacobian linear with respect to $u$ and $v$  vanish  generically at
points in the $(x,y)$ plane. Of course, besides this cancelation,
there are constraints on the sign of $a$ and of the quadratic form
$b u^2 + c v^2 + 2d u v$ to make the singularity appear as time
increases. For instance, if $a$ is positive, the quadratic form $b
u^2 + c v^2 + 2d u v$ must be negative definite, which requires $c$
and $b$ negative and $d^2 - ab$ negative.

The functions $F(u,v)$ and $G(u,v)$ have to be expanded in Taylor
series at least to third-order to yield the quadratic term $b u^2 +
c v^2 + 2d u v$ in the Jacobian. This depends on two sets of eight
coefficients. Fortunately this calculation is greatly simplified by
limiting oneself to the quantities at leading order near $t =0$. The
Jacobian at leading order depends on the
 linear terms in the TSE of $F$ and $G$ near $u = v =0$.
 Let us note that  coordinates $x$ and $y$ do not have to be taken in directions orthogonal to each
  other,
  because the Jacobian matrix is reduced to its diagonal form by a general change of coordinates that
  is not
   necessarily  a rotation, the Jacobian matrix having no reason to be real symmetric in general.
At $t =0$ the Jacobian matrix has a zero eigenvalue, that can always
be associated with an
 eigenvector pointing in the $x$ direction. Therefore the coefficients of the TSE of
 $F(u,v)$
 and $G(u,v)$ have to satisfy the condition that $\frac{\partial F}{\partial u} = 0$, $\frac{\partial F}
 {\partial v} = 0$ and $\frac{\partial G}{\partial u} = 0$ at $ u = v = 0$. This yields that, near the
 singularity, one has at leading order
\begin{equation}
 \left \{ \begin{array}{l}
 F(u,v) = b' v^2 + d' u^3\\
  G(u,v)=c'v
 \mathrm{,}
 \label{fg}
\end{array}
\right.
\end{equation}
with $b', c' $ and $d'$ coefficients depending on the initial
  conditions for the original equation, which are related to the constant
$a,b$ of expansion (\ref{eq:9}) by the relation $a=c'$,
  $b=3c'd'$. Therefore at leading order, the second
  equation (\ref{eq:FG}) writes

\begin{equation}
 y = c' v
 \mathrm{,}
\label{eq:y=v}
\end{equation}
 because one can neglect
   the term $v t$ as compared to $c'v$ (because $t$ is a priori much smaller
  than the constant $c'$, assumed non-zero). The first equation (\ref{eq:FG})
becomes:

\begin{equation}
x - \frac {b'}{c'^{2}} y^2 - u t = d' u^3 \mathrm{.}
 \label{eq:8c}
\end{equation}

 This equation is not yet the one we are aiming for, because it
describes a singularity all along the parabola of Cartesian equation
$x - \frac{b'}{c'^{2}} y^2  = 0$ at time $0$. Actually we omitted a
term of order $u v^2$ in the TSE of $F(u,v)$. Let us thus write
$F(u,v) = b' v^2 + d' u^3 + g' u v^2$. The term $g' u v^2$
 ultimately yields
  a term of order $ y^2 u$ in the equation for $u$. Contrary to the term
 $ - \frac{b'}{c'^{2}} y^2$ just
 considered, this one changes the singularity completely in the sense that, with this term,
 the singularity occurs first at a point in the $(x,y)$ plane and later spreads in the $y$ direction.
 Thus, after some elementary rescaling, equation (\ref{eq:4a}) for the 1D case is changed into:

\begin{equation}
0  =  u^3 + x - \beta y^2- u (t - y^2) \label{eq:8d}
\end{equation}
which is the main result of this section.

 Note, we neglected the term $m'u^2 v$ in the TSE of $F(u,v)$, although it is of the same order of
magnitude, $|t|^{3/2}$, as the one we kept. This is because this
term yields a term like $ u^2 y$ in the equation for $u$, which can
be eliminated by redefining $u$ as
 $ u + m"y$, with $m"=\frac{m'}{3}$, a constant.
Also, first, note that for actual overturning waves in shallow
water, at least at leading
 order, the velocity component $u$ is proportional to the surface
 elevation (see section 4 where simple waves are defined for this case).
From now on, let us consider $u$
 as the surface elevation, until we can further define the elevation
 origin. Second, note that in  equation (\ref{eq:8d}), the term $- \beta y^2$ changes the
support of the singularity from a straight line to a bent line.
 Additionally, except for this term, all other terms in
equation (\ref{eq:8d}) are of order $|t|^{3/2}$ near the
singularity, with
\begin{equation}
y\sim |t|^{1/2},
 u \sim |t|^{1/2} , x \sim |t|^{3/2}
 \mathrm{.}
 \label{eq:scaling}
\end{equation}

 Now, for a given value of $t$, the extension of the  overhanging
 region of the 3D surface $u(x,y,t)$ may be visualized by  the set of points where the tangent plane
 is vertical; this set of points will be referred to as the apparent contour $\it{C_{v}}$.
 Differentiating equation (\ref{eq:8d}), we get
  the relation
 $(3u^{2}+y^{2}-t)du + dx +(2u-2\beta)dy =0$. Along $\it{C_{v}}$, the condition of a vertical
 tangent, i.e., infinite values of partial derivatives ($ \frac{\partial u}{\partial
 x}$, $\frac{\partial u}{\partial
 y}$), requires that the coefficient of the $du$ term vanish. Introducing the relation
 $3u^{2}+y^{2}-t=0$ into equation (\ref{eq:8d}),
  yields the equation for the apparent contour,
\begin{equation}
x=\beta y^{2}\pm \frac{2}{3^{3/2}}(t-y^{2})^{3/2}
\mathrm{.}
\label{eq:proj}
\end{equation}

\begin{figure}[ht]
\centerline{\includegraphics[scale=1.2]{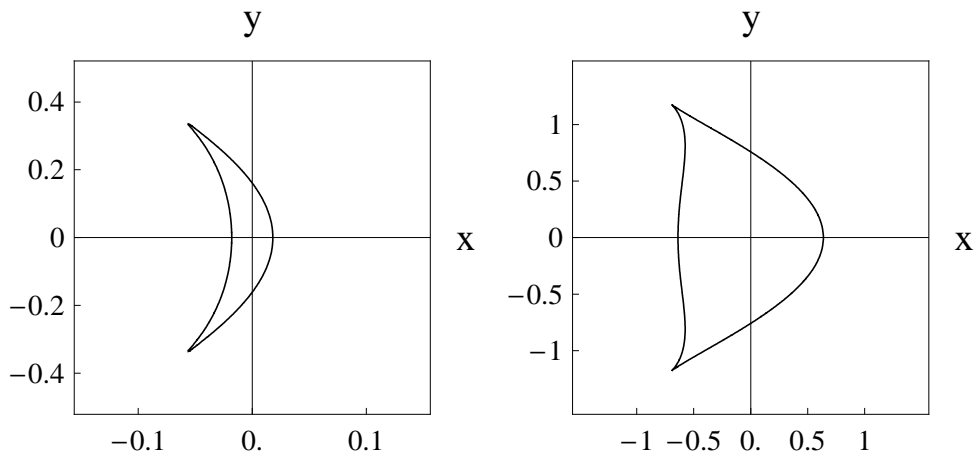}}
 \caption{Boundary of the overhanging region or apparent contour
 $\it{C_{v}}$, given by equation   (\ref{eq:8d}), with $\beta=-0.5$.
  Leftward figure : just after the
 beginning of the overturning process at $t=0.13$; rightward figure :  for
 $t=1.4$, which is actually beyond
 the  range of validity of the  'generic equation'}.
\end{figure}

 Equation (\ref{eq:proj}) is plotted in Fig. (1), and shows the
boundary of the overhanging domain projected on the horizontal
plane, at time $t$ after the inception of the singularity. This
boundary spreads along the parabolic curve of Cartesian equation
$x=\beta y^{2}$. The width of the crescent shaped boundary expands
as $x \sim |t|^{3/2}$, while its length expands as $y\sim
|t|^{1/2}$.

These analytical predictions are compared in the next section to
results of numerical simulations of wave-breaking, derived from
those reported in \cite{GG}.

\subsection{Numerical simulation of wave-breaking}
\label{subsec:numericalwave}

Euler equations with fully nonlinear free surface conditions have
recently been solved numerically in 3D,  to simulate the early
stages of solitary wave-breaking induced by changes in topography in
shallow water, namely a curved sloping ridge \cite{GG}. Note that
the word "solitary" refers here to a single, isolated wave, not to a
soliton-like wave propagating without changing shape. The flow in
the initial solitary wave being irrotational, Kelvin theorem implies
that this will be the case for any positive value of $ t$. Hence,
the problem is described by fully nonlinear potential flow
equations, which are solved in the model with a high-order boundary
element method, combined with an explicit time-integration method,
expressed in a mixed Euler-Lagrangian formulation.  In fact, earlier
comparisons of 2D numerical results with laboratory experiments have
 shown that the full potential theory predicts well
wave overturning in deep water, as well as wave shoaling and
overturning over slopes \cite{Grilli}. Recent contributions to 3D
simulations of nonlinear and overturning waves are reviewed in
\cite{GG}, where new results for the velocity and acceleration
fields during wave overturning are also computed, both on the
surface and within the flow.

\begin{figure}[ht]
    (a) \centerline{\includegraphics[scale=0.4]{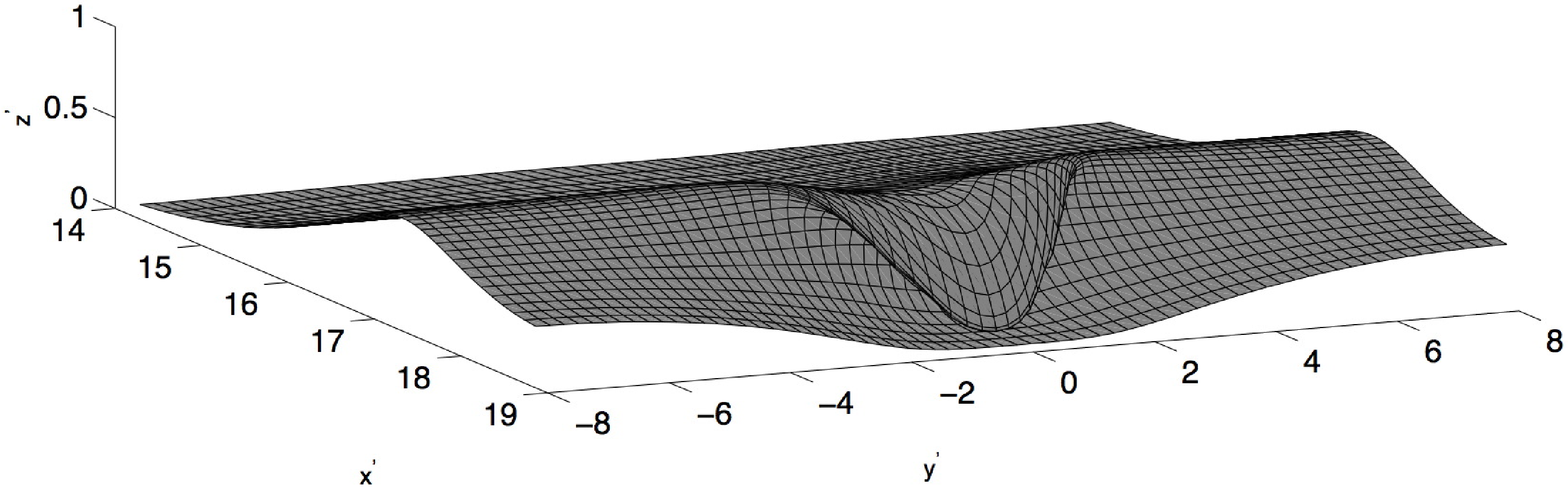}} \\
    (b) \centerline{\includegraphics[scale=0.6]{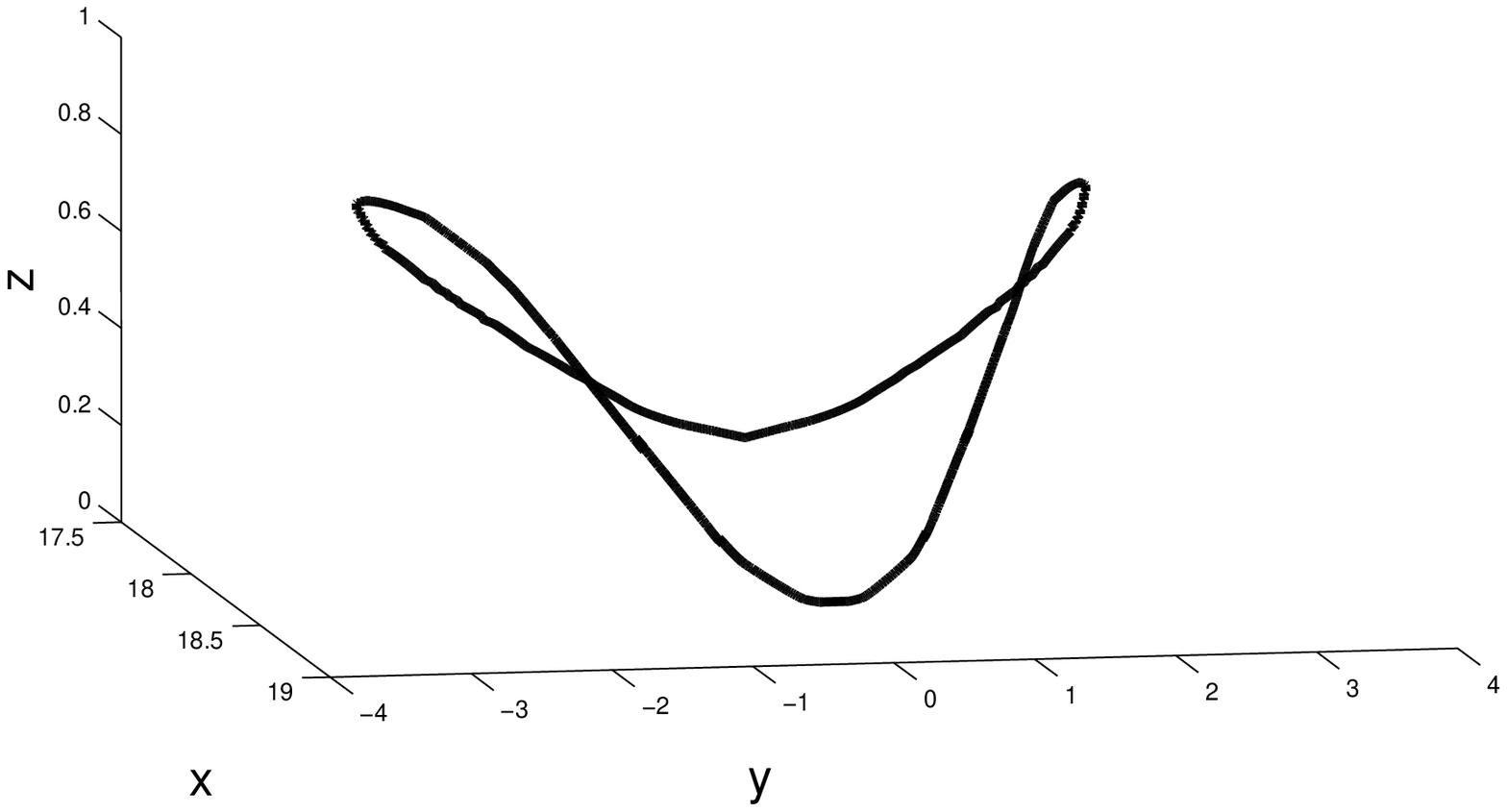}}
    \caption{ Numerical results at $t=1.378$ (based on  \cite{GG}): (a) 3D wave profile; and (b) apparent
    contour}.
    \end{figure}

 Some of the numerical results reported in \cite {GG} are revisited
below, and compared to the theoretical predictions of section
\ref{subsec:wavebreakin2D}.

An idealized geometry was selected to simulate 3D breaking-waves on
beaches, which is shown in figure 2 of ref.\cite {GG}. The wave
propagates towards the positive $x$ direction, where the water depth
$z=-h_{0}$ is constant in the first part of the tank, then a sloping
ridge starts at $x=5.22$ with a 1/15 slope in the middle
cross-section and a transverse modulation of the form
$sech^{2}(ky)$, with $k=0.25$ in the present case. (the
non-dimensional variables $(x,y)$ being scaled with $h_{0}$ and time
with $\sqrt{h_{0}/g}$). On such a bathymetric lens, that causes
directional wave focusing, the wave first overturns on the axis, at
$y=0$ and then breaking propagates laterally in both $\pm y$
directions. 3D wave profiles are shown in Figs. $5c, 6c$, and $20$,
of reference \cite{GG}, for time values after overturning $t=1.036,
1.252$ and $1.378 $, respectively ($t= t'-t'_{B}$ in \cite{GG}'s
notations).

\begin{figure}[ht]
    \begin{center}
    \includegraphics[scale=0.6]{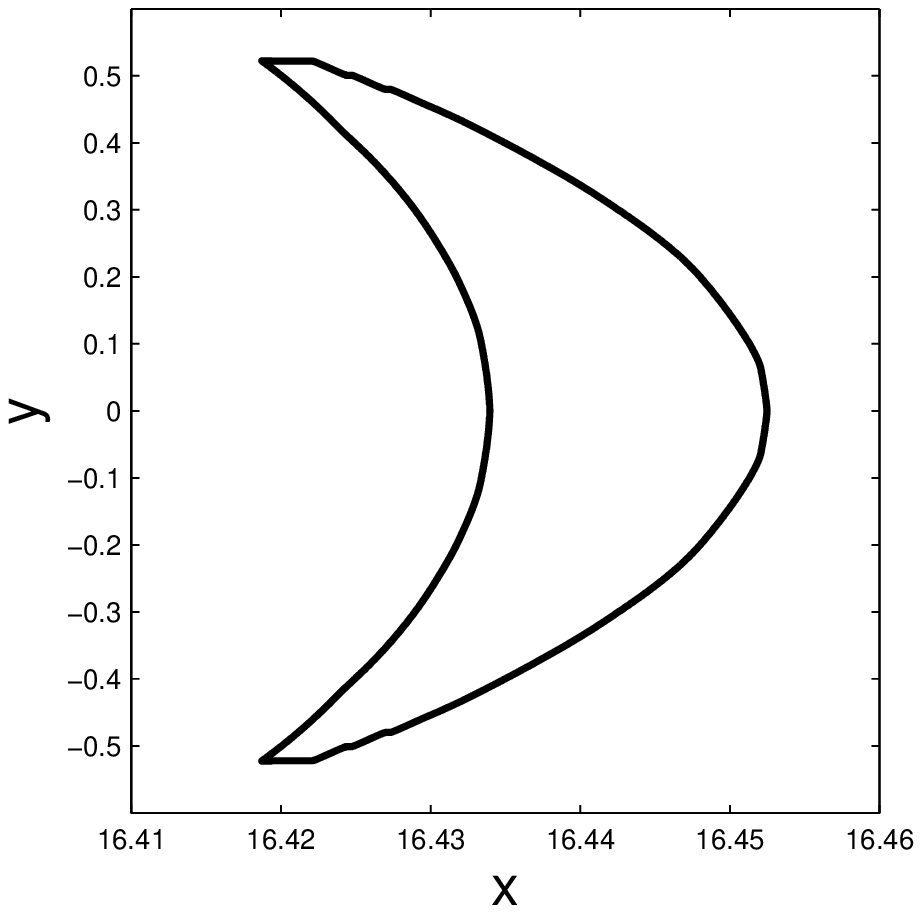}
    \includegraphics[scale=0.6]{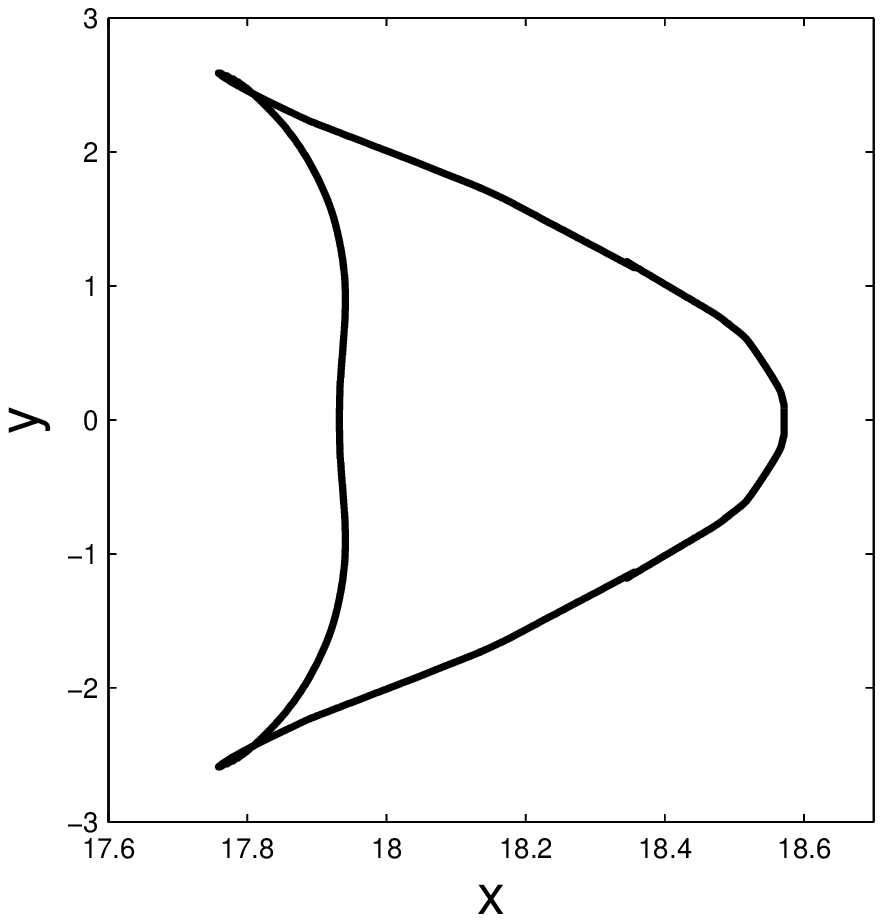}
    \end{center}
    \caption{Numerical results (based on  \cite{GG}): Projection of the apparent contour at
    time (left) $t=0.13$, (right) $t=1.378$}.
    \end{figure}

 In Figure 2a, we reproduce the 3D wave profile at time $t=1.378$
just before the plunging jet reaches the underlying surface of the
wave, and show in Figure 2b the apparent contour of the
corresponding overhanging region. Figs. 3 displays  apparent
contours projected on the horizontal plane, both for this and an
earlier time case, $t=0.13$. Both of these apparent contours are in
good qualitative agreement with the theoretical predictions of
equations (\ref{eq:8d})-(\ref{eq:proj}), shown in Figure 1 for the
same two values of $t$. Hence, even though the scaling laws were
derived above for small values of ($t,u,x,y$)  we observe that the
'generic equation' derived in the previous subsection remains
correct beyond this range of validity. Indeed the  apparent contours
for the latter time are very similar, see Figs. 2b and 3 (right), while the corresponding values
of ($t,u,x,y$) become larger than unity.

\begin{figure}[ht]
    \centerline{\includegraphics[scale=0.6]{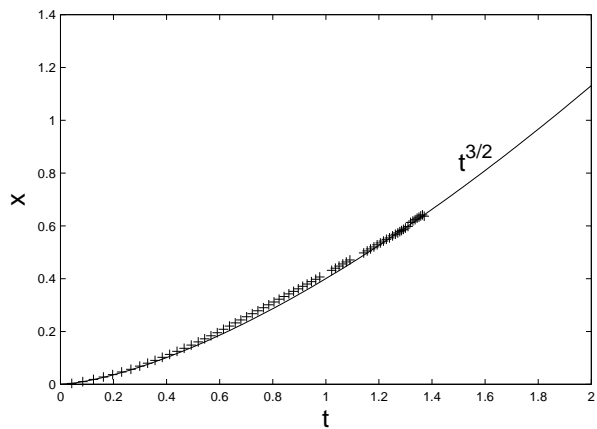}}
    \caption{Numerical results (+) (based on  \cite{GG}), compared to analytic prediction, equation (\ref{eq:scaling}): Growth
    of the overhanging region along the propagation direction $x$, as a function of  $t= t'-t'_{B}$}.
    \end{figure}

\begin{figure}[ht]
    \centerline{\includegraphics[scale=0.6]{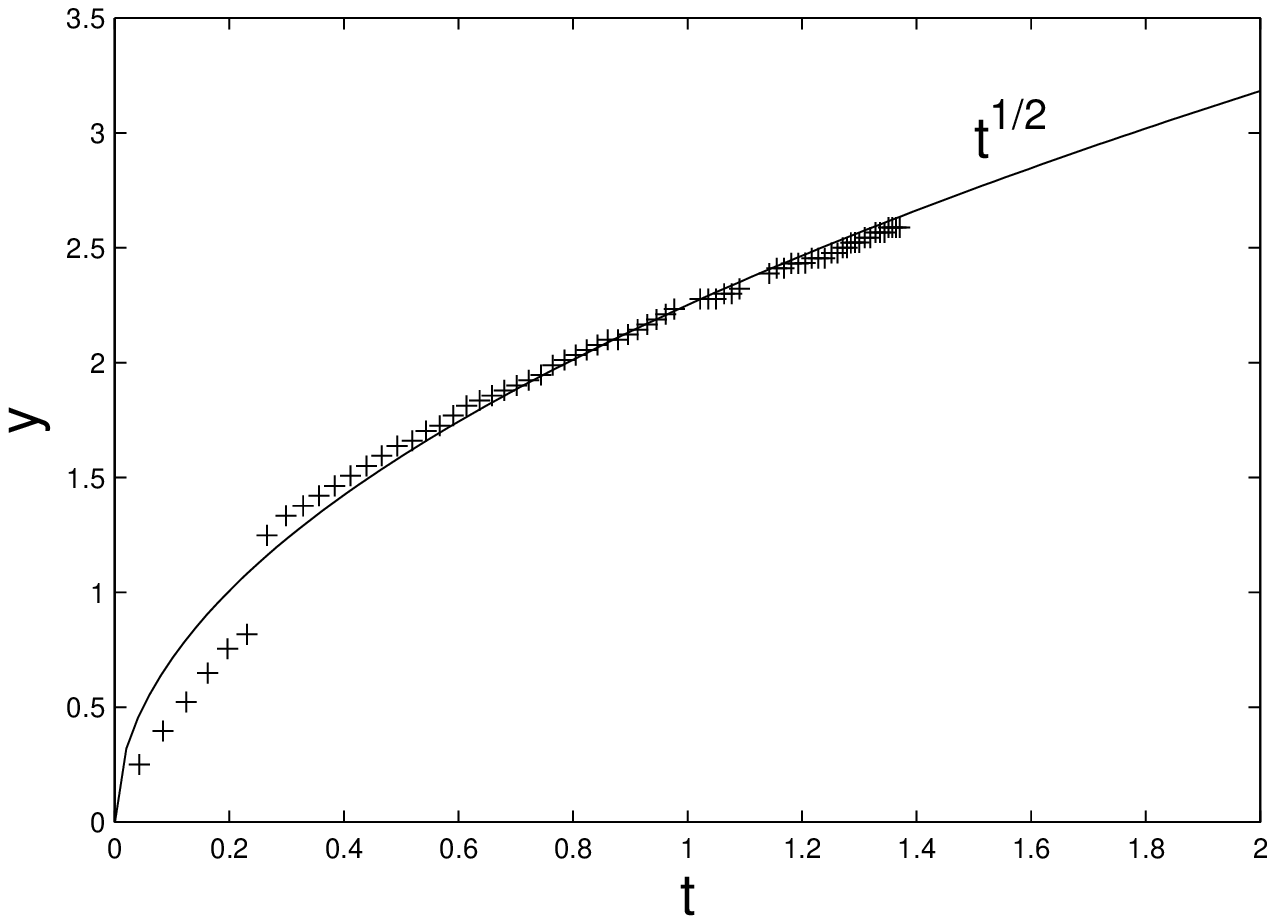}}
    \caption{ Numerical results (+) (based on  \cite{GG})
    compared to analytic prediction, equation (\ref{eq:scaling}): Growth of the overhanging region in the  direction $y$ transverse
     to propagation
    as a function of  $t= t'-t'_{B}$}.
    \end{figure}

Let us now test the scaling laws for the longitudinal and transverse
expansions in time of the singularity (defined as $h(x,y,t)$ being
multivalued in the overhanging region), representing the overturning
wave. Thus, Figs. 4 and 5 show numerical results for the spatial
expansion of the apparent contour along $x$ and $y$, respectively,
as a function of the time increase from the beginning of the
overturning. The two curves display a good
 fit with the power laws $t^{3/2}$ and $t^{1/2}$,
respectively, as predicted by our analysis in section
\ref{subsec:wavebreakin2D}. The discontinuity near time $t=0.23$ in
Figure 5 is likely a numerical artifact due to the difficulty of
accurately defining the apparent contour along the $y$ direction for
small values of $t$. In summary the predictions of the `generic
solution'(\ref{eq:proj}), proposed above, are well confirmed by the
simulations.

While these simple scaling laws have been confirmed, the asymmetry
of the actual wave profile, due to the appearance of a plunging jet,
cannot be described by the simple cubic term in the ($F,G$)
expansion. In fact, the plunging jet can be viewed as a late
evolution of a Rayleigh-Taylor instability, which occurs when a
heavy liquid is placed over a lighter one.  This phenomenon was
recently studied in \cite{Clavin}, where a finger in free fall was
also observed in the late (and highly nonlinear) stage of the
Rayleigh-Taylor instability. The 2D analysis of \cite{Clavin}
assumed a jet falling vertically (along $z$ in our notations)
without horizontal velocity (along $x$). We will extend this work to
estimate further properties of the overturning jet, by including an
initial longitudinal velocity of the jet, $u_{0}$. Thus, in the
overturning region, the tip of the jet follows a ballistic motion,
with $z\sim{-\frac{1}{2}gt^{2}}$, and a quasi-constant horizontal
velocity (see, e.g., Figure 12a of \cite{GG}). Scaling both time and
position, as in \cite{Clavin}, by factors $\sqrt{gk}$ and $k$
respectively (where $k$ is the wave number of the implemented sine
perturbation of the interface) and changing their notations $y, v$
into $-z,w$ to
 avoid confusion with the above notations, the velocity field in the tip
 region is approximated by

 \begin{equation}
 \left \{ \begin{array}{l}
 u\sim{u_{0} }\\
 w\sim{-t}\sim{-\sqrt{2z}}
 \mathrm{.}
 \label{eq:uw}
\end{array}
\right.
\end{equation}
Now, the kinetic equation for the interface $\alpha(x,t)$ reads
\begin{equation}
\frac{\partial \alpha}{\partial t} + u_{0} \frac{\partial
\alpha}{\partial x} = -\sqrt{2\alpha}
 \mathrm{.}
 \label{eq:alfa}
\end{equation}
Defining $\alpha(x,t)=-t[t/2-\gamma(x,t)]$, we obtain after
linearization
\begin{equation}
\frac{\partial \gamma}{\partial t} + u_{0} \frac{\partial
\gamma}{\partial x} = 0
 \mathrm{.}
 \label{eq:gamma}
\end{equation}
which can be solved in the form, $\gamma(x,t)=\theta (x-u_{0}t)$.
The curvature of the plunging tip is  then defined as
$\kappa=\frac{\partial^{2} \alpha}{\partial t^{2}}|_{x =u_{0}t}$.
This yields the scaling relationship for the asymptotic evolution of
the curvature
\begin{equation}
\kappa\sim |t|\theta _{0}^{"} \mathrm{.}
 \label{eq:clavin2}
\end{equation}
[Note that, without the horizontal velocity, one finds $\kappa\sim
|t|^3\theta_{0}^{"}$.]

\begin{figure}[ht]
\centerline{\includegraphics[scale=0.6]{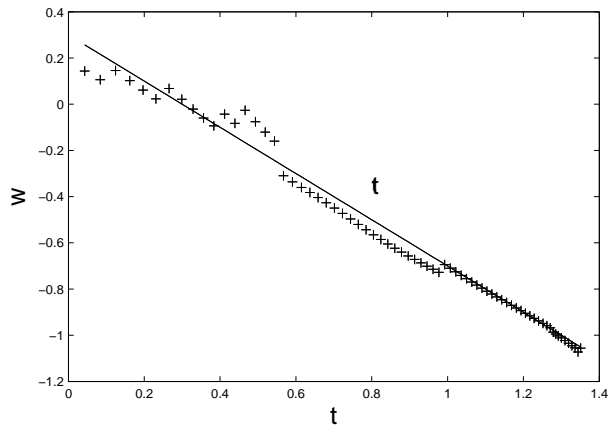}} \caption{
Numerical result (+) (based on \cite{GG}), compared to analytic
prediction, second equation (\ref{eq:uw}): Growth of the vertical
velocity at the jet end, at $y=0$.}.
\end{figure}

\begin{figure}[ht]
\centerline{\includegraphics[scale=0.6]{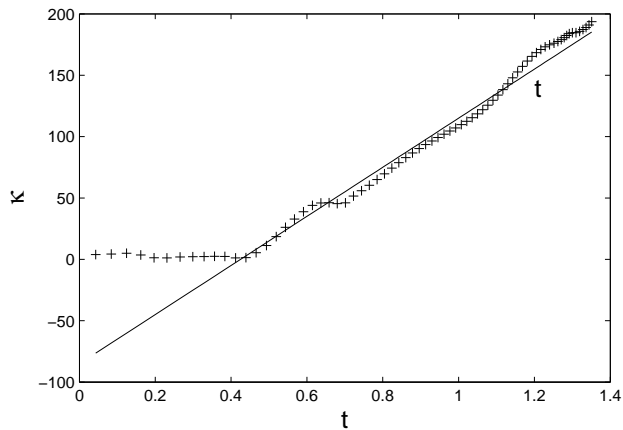}}
\caption{ Numerical result (+) (based on \cite{GG}), compared to
analytic prediction, equation (\ref{eq:clavin2}):  Growth of the
curvature at the jet end, in the plane $y=0$.}.
\end{figure}

We now posit that the predictions (\ref{eq:uw}) and
(\ref{eq:clavin2}) of the asymptotic linear evolution of $w$ and
$\kappa$ with time, are also valid in the 3D case. Indeed the first
one simply reflects the ballistic motion, and the second one remains
valid when the curvature of the wavefront along the crest is taken
into account (this being time independent at leading order, see
equation (\ref{eq:proj})). Figs. 6 and 7 confirm that the
relationships  (\ref{eq:uw}) and (\ref{eq:clavin2})  agree well with
results of 3D simulations.

\section{Wave-breaking in shallow water}
\label{sec:wavebreakinginshallow}

 The analysis we presented before
does not directly apply  to actual waves, which are not physically
described  by the pressure-less fluid equations used so far.
However, we will show below that, for surface waves propagating in
shallow water, the
 equations governing wave motion reduce to
the equations of a pressure-less fluid near the singularity.

Weakly nonlinear and dispersive inviscid long waves are well
described by Boussinesq equations, when $A/h_{0} \sim (kh_{0})^{2}
\ll 1$ \cite{Mei}, with $A$ a characteristic wave amplitude,
$k=2\pi/\lambda$, the wavenumber with $\lambda$ the wavelength, and
$h_{0}$ the water depth from mean water level. More recently,
extended Boussinesq models (BM) have been proposed that apply to
strongly nonlinear waves, with $A/h_{0} = {\cal O}(1)$, and have
been shown to stay valid close to the breaking point \cite{Wei}. In
BMs, the fundamental quantities are the total fluid depth,
$h(x,y,t)$ (i.e., from free surface to bottom)  and the two
Cartesian components of the
 horizontal fluid velocity, $u(x,y,t)$ and $v(x,y,t)$.
The BM equations of motion express the conservation of mass and of
horizontal momentum. Assuming fully nonlinear non-dispersive long
waves, BM equations yield the "nonlinear shallow water wave"
equations \cite{Mei}:
\begin{equation}
 \left \{ \begin{array}{l}
\frac{\partial h}{\partial t} + \frac{\partial (u h)}{\partial x} +  \frac{\partial (v h) }{\partial y}= 0\\
\frac{\partial u}{\partial t} + u \frac{\partial u }{\partial x} + v
\frac{\partial u}{\partial y} + g \frac{\partial h}{\partial x} =
0\\
\frac{\partial v}{\partial t} + u \frac{\partial v }{\partial x} + v
\frac{\partial v}{\partial y} + g \frac{\partial h}{\partial y} = 0
\mathrm{,} \label{eq:Bouss}
\end{array}
\right.
\end{equation}
where $g$ is the positive acceleration of gravity. Whenever all
functions are independent  of $y$, this set of equations admits
nontrivial simple 1D wave solutions. We shall see that similar
results also exist in 2D.

\subsection{One-dimensional case}
When the solution only depends on $x$, we can define $h = h_0 +
\tilde{h}(x,t)$ and, if we further assume that  $h_0$ is much larger
than the free surface perturbation $\tilde{h}(x,t)$, one may
neglect\cite{Mei} terms of order $\tilde{h}(x,t) u$  with respect to
terms of order $h_0 u(x,t)$ in the BMs. This yields:
\begin{equation}
 \left \{ \begin{array}{l}
\frac{\partial \tilde{h}}{\partial t} + u \frac{\partial
\tilde{h}}{\partial x} + h_0 \frac{\partial u}{\partial x} = 0\\
\frac{\partial u}{\partial t} + u \frac{\partial u }{\partial x} + g
\frac{\partial \tilde{h}}{\partial x} = 0
 \mathrm{.} \label{eq:11}
\end{array}
\right.
\end{equation}
Note that, although $\tilde{h}$ has been neglected with respect to
$h_0$, one has kept the term $u \frac{\partial \tilde{h}}{\partial
x}$ in the first equation, because  a priori it is not smaller than
any other term written explicitly in the second equation
(\ref{eq:11}).

Let us  now scale the lengths with $h_{0}$ and time with
$\sqrt{h_{0}/g}$, that amounts to divide the velocity by  the long
wave celerity $C = \sqrt{g h_0}$. In terms of scaled variables, the
equations for the new functions $ \tilde{h}$ and $u$ thus become:
\begin{equation}
\left \{ \begin{array}{l}
 \frac{\partial \tilde{h}}{\partial t} + u
\frac{\partial \tilde{h}}{\partial x} + \frac{\partial u}{\partial
x} = 0\\
\frac{\partial u}{\partial t} + u \frac{\partial u }{\partial x} +
\frac{\partial \tilde{h}}{\partial x} = 0
 \mathrm{.} \label{eq:11c}
\end{array}
\right.
\end{equation}

  If one assumes a \emph{simple wave} solution of Equations \ref{eq:11c}, namely that $ u =
\tilde{h} $, the two equations  become identical and yield
 the familiar (so-called) Burgers equation
(\ref{eq:1}), when replacing $u$ (resp. $\tilde{h}$) by $(u+1)$
(resp.$(\tilde{h}+1)$). This shows that the singularity in these
equations is
 the same one as found before. This is of course not a new result, because of the possibility of
 solving the full set of equations (\ref{eq:Bouss}) by Riemann's method
 with $v=0$ and no dependence on $y$.

\subsection{ Waves in two dimensions}
 Let us now use the full 2D equations (\ref{eq:Bouss}) and thus define $h = h_0
+ \tilde{h}(x,y,t)$, with $\tilde{h}$ still being small, compared to
$h_0$. Using the same scaling as above, the equations
(\ref{eq:Bouss}) for the mass and momentum conservation write in
terms of scaled variables:
\begin{equation}
\left \{ \begin{array}{l}
 \frac{\partial \tilde{h}}{\partial t} +
\left( \frac{\partial u}{\partial x} + \frac{\partial v}{\partial
y}\right) + u \frac{\partial \tilde{h}}{\partial x} + v
\frac{\partial \tilde{h}}{\partial y}= 0 \\
\frac{\partial u}{\partial t} + u \frac{\partial u }{\partial x} + v
\frac{\partial u}{\partial y} + \frac{\partial \tilde{h}}{\partial
x} = 0\\
\frac{\partial v}{\partial t} + u \frac{\partial v }{\partial x} + v
\frac{\partial v}{\partial y} + \frac{\partial\tilde{h}}{\partial y}
= 0 \mathrm{.}
 \label{eq:14}
\end{array}
\right.
\end{equation}

To check whether these equations approach a  pressure-less
Bernouilli equation  near the wave-breaking singularity, one assumes
that the derivative
 $\frac{\partial v}{\partial y}$ is non-singular near this singularity, although $\frac{\partial u}{\partial x}$
 diverges (which is true for simple waves, see below).
 Therefore one can neglect the former with respect to the latter.
 Hence, near the singularity,
 the first equation (\ref{eq:14}) may be replaced by
\begin{equation}
\frac{\partial \tilde{h}}{\partial t} + u \frac{\partial
\tilde{h}}{\partial x} + v \frac{\partial \tilde{h}}{\partial y}+
\frac{\partial u}{\partial x} = 0 \mathrm{.} \label{eq:15a}
\end{equation}

Now, if one assumes a simple wave solution, as in the 1D case,
namely that $u =\tilde{h}$,  Equation (\ref{eq:15a}) becomes
identical to the second equation (\ref{eq:14}). The first and second
equations thus coalesce into $\frac{\partial u}{\partial t} + u
\frac{\partial u}{\partial x} + v \frac{\partial u}{\partial y}+
\frac{\partial u}{\partial x} = 0 \mathrm{.}$ Finally defining
$\tilde{u}=u+1$, the set of equations (\ref{eq:14}) writes,

\begin{equation}
\left \{ \begin{array}{l} \frac{\partial \tilde{u}}{\partial t} +
\tilde{u} \frac{\partial \tilde{u}}{\partial x} + v \frac{\partial
\tilde{u}}{\partial y}=0\\
\frac{\partial v}{\partial t} + \tilde{u}\frac{\partial v }{\partial
x} + v \frac{\partial v}{\partial y} +(
\frac{\partial\tilde{u}}{\partial y} - \frac{\partial v}{\partial
x})=0 \mathrm{,} \label{eq:16}
\end{array}
\right.
\end{equation}
 which is identical to the pressure-less fluid system (\ref{eq:6}),
because the expression $ (\frac{\partial\tilde{u}}{\partial y} -
\frac{\partial v}{\partial x})$ vanishes for irrotational fluids.

This shows that, at least at leading order, wave-breaking scaling
laws are identical for a pressure-less fluid and for the nonlinear
shallow water wave equations.

This brings an interesting point: it seems that the widening of the
wave-breaking domain in the $y$ direction, like the square root of
time,  is 'universal' for non-linear non-dispersive waves. Other
kinds of waves are known to break, like  ocean waves on deep water
(with, practically infinite depth). Does their spreading in the $y$
direction follow the same square root law ? The answer to this
question depends
 in particular on how precisely one can identify the time for the
 onset of the breaking process.
This deserves further investigations.

\section{Discussion}

\subsection {Regularisation}

 The regularization of the singularity appearing in solutions of the
inviscid Burgers equation has been extensively studied.  One such
regularization scheme relies on an assumed balance of nonlinear
 and  dispersive effects over a non-vanishing fluid depth. This
yields the well-known Korteweg-de Vries (KdV) equation in 1D. This
equation has been  very thoroughly studied and can be shown  to be a
form of the standard Boussinesq equations when assuming wave
propagation occurs in a forward direction only \cite{Mei}. Hence, as
for the standard weakly nonlinear and dispersive BM mentioned above,
KdV equation requires  that the two physically unrelated processes
of nonlinearity and dispersion be of the same order. Hence, the
range of applicability of this equation
 to water waves is quite narrow.

 A second way  to avoid the singularity consists in adding a small
dissipative term  to the inviscid Burgers equation (\ref{eq:1}) to
make it ressemble the Navier-Stokes equation. This yields the
so-called Burgers equation,
 \begin{equation}
\frac{\partial u}{\partial t} + u \frac{\partial u}{\partial x} =
\epsilon \frac{\partial^2 u}{\partial x^2} \mathrm{,} \label{eq:5}
\end{equation}
 with the coefficient $\epsilon$ being small and positive. With a
non-zero $\epsilon$ value, the solution of equation (\ref{eq:5}) is
not multi-valued anymore (the equation is regularized) and one has
to replace in the above analysis the multi-valued $U_{1}$ by a
function with a jump at $\zeta =0$, which represents a shock wave
for this model. By symmetry, the shock discontinuity is at $\zeta =
0$. There,  $U_{1}$ jumps from $+1$, for slightly negative values of
$\zeta$, to $(-1)$, for slightly positive values of $\zeta$. In a
small interval near $\zeta = 0$, of width of order $(\epsilon
t)^{1/2}$, the solution goes continuously from $+1$ to $(-1)$ across
the shock front. Moreover, just at the time of the discontinuity,
there is also a time interval where the added diffusion term is
nowhere negligible. This time interval is such that the thickness of
the self-similar region in $x$ is of the same order as the typical
length scale associated with diffusion. The former decreases to zero
as $t$ tends to zero, like $|t|^{2/3}$, and the latter like
$(\epsilon t)^{1/2}$. The cross-over time is such that the two
length scales are of the same order of magnitude, which happens for
$t\sim \epsilon^3$. For time scales of this order or shorter, the
self-similar solution is not valid anymore and another limit has to
be taken. Real wave-breaking, however, cannot be described  by this
regularization, because it implies an overturning of the surface
that is not represented by the viscous-like term on the right-hand
side of equation (\ref{eq:5}).

\subsection{Geometrical meaning of the singularity or where is the true singularity?}
\label{subsec:geometry}

Before proceeding further,  it is worthwhile reconsidering the
question of the singularity in  light of a geometrical approach.
This will yield a more precise definition of what we mean by
wave-breaking. Let us get back to the 'generic' solution
(\ref{eq:4a}) of the pressure-less fluid equation (\ref{eq:1}). In
the simple wave case, this solution takes a simple geometrical
meaning in which the two quantities $u$ (velocity component along
the propagation direction) and $h$ (elevation of the fluid surface)
are proportional to each other, up to irrelevant multiplicative
constants. Then the 1D solution of,
\begin{equation}
0  =  h^3 + x - h t \mathrm{} \label{eq:4c}
\end{equation}
can be seen as describing the occurrence of an apparent contour of a
curve drawn on a plane, as it is continuously deformed, the
deformation parameter being $t$. Such an apparent contour is
defined (see section \ref{sec:from1D2D}) as the set of points where the tangent is parallel to the
direction of observation, and the time origin  $t =0$ is the instant
where all points collapse into a single one. It is worthwhile
noting that this definition depends upon the direction of observation.
Indeed in 1D, at $ t =0$, the local Cartesian equation of the curve
is $h^3 + x =0$, $h$ being the abscissa and $x$ the ordinate. As
time changes, the curve is deformed and generically takes a non zero
slope at the origin. The slope increases linearly with time, while
the two extrema of the curve $x(h,t)$ move apart from each other, as $\Delta x \sim
 t^{3/2}$. Therefore the equation  (\ref{eq:4c}) describes the
local shape of a line, when it acquires a
   contour at $t =0$. This contour is fundamentally linked to the direction of observation and would
   vanish if this direction were tilted in the direction parallel to the
   slope, at the inflection
   point. While it seems natural to choose the  vertical axis as the observation direction in real life,
    we must note
   that the description of wave-breaking as the appearance of a contour is
   ambiguous.

 To get round this difficulty let us note that wave-breaking could be defined either by the occurrence of
  overturning (i.e., an overhanging region on the surface),
 or by the impact and intersection of the plunging jet with the surface underneath. Casual observation indicates
 that the two phenomena are very strongly linked in the sense that overturning usually leads first
 to plunging  and then to impact with the underlying surface. However things
 are not generally so simple, especially because the equations of motion are reversible (viscosity is
 negligible and thus neglected for this typically large Reynolds
  number flow). Hence, under time-reversal symmetry, a plunging jet
  would first retract onto itself and later
  return  to a disappearing apparent contour.

  Such a reversible event, namely the
    possibility of reversible overturning (i.e., the overhanging
    region spontaneously retracting  instead of becoming a plunging jet), could possibly occur. This could happen, for instance, for
       a wave propagating over a decreasing depth, that would
      initiate   overturning,
      then entering a region of rapidly increasing depth, which could make the
      overhanging region gradually disappear, provided its
      development has not proceeded too far in time. Another case could be
      that of a surface pressure (i.e, caused by wind) gradually
      increasing and  `pushing' against the nascent overturning
      region. Therefore an upward moving retracting jet is certainly among the solutions of the full dynamical problem
 and  would lead  quite naturally to define
     wave-breaking as the occurrence of self-penetration of the surface, a definition free of
    arbitrariness because, contrary to anything related to the apparent contour, it does not rely on an
    arbitrary choice of the direction of observation.

      Seen from another perspective,  wave-breaking now defined  by the
      self-penetration of the surface by the plunging jet follows,
       but not always, from overturning of the surface. This raises also the very interesting issue of
       describing the post-penetration flow and surface intersection behavior (discussed below).
       Indeed regularization by capillary phenomena,
    viscosity, etc\ldots may make this a well posed problem. Nevertheless we
    posit that the problem of
    impact between the plunging jet and the surface underneath it can be formulated in almost the same
    manner as the problem of propagation of nonlinear waves. This is
    discussed in the next subsection.

        \subsection{Post collision dynamics}
        \label{subsec:postcoll}

 Classically the dynamics of nonlinear waves in the inviscid limit can
be formulated as a system of two coupled
 first-order (in time) equations for the value of the velocity potential on the free surface and for the
  position of an arbitrary point on this surface. Let us briefly summarize
  the bases for this approach, in the usual
  case without collision of fluid interfaces. Knowing the value of the velocity potential on the free surface
   and, given the Neumann condition on the bottom, one can in principle solve the Dirichlet-like
   Boundary Integral Equation (BIE) problem for
   the velocity potential everywhere \cite{GG,GGD}. This yields in particular the velocity normal to the free surface that
   is used to advance this surface in time. The value of the velocity potential on the surface is
   then similarly advanced by time integrating  the Bernoulli equation at every
   free surface point.
    The major advantage of this BIE method is
    that, in numerical studies, one can transform the problem of solving Laplace or Poisson's equation
    over the entire domain into
     an integral equation on the surface, solvable by analytic function theory in 2D
     and by Green's function method in both 2D and 3D
     \cite{Grilli,GGD}, short-cutting the computationally
     complex problem of drawing a
     smooth curve or a surface embedded in the geometrical space.

 To summarize, one starts from the unsteady Bernoulli equation for the velocity potential $\Phi({\bf{r}},t)$:
 \begin{equation}
\frac{\partial \Phi}{\partial t} +  \frac{1}{2}\left(\nabla \Phi\right)^2 + g z  +  \frac{p}{\rho} = 0
\mathrm{}
\label{eq:Be1}
\end{equation}
with $z$  the height, $\rho$ the mass density of the fluid and $p$ the pressure. This equation is expressed
on the free surface of the fluid, located  at
 $ z = h({\bf{r}}_{H},t)$, where ${\bf{r}}_{H}$ is the set of horizontal coordinates.
 Note that if there is overturning, the function $h({\bf{r}}_{H},t)$ is not
 single-valued everywhere. Because the free surface is in contact with air, the atmospheric
 pressure is applied everywhere on the surface, whose variation can be
 neglected. Therefore, we can set  $p=p_{a}=0$ at $ z = h$ and equation (\ref{eq:Be1}) becomes:
 \begin{equation}
\frac{\partial \Phi}{\partial t} =  - \frac{1}{2}\left(\nabla
\Phi\right)^2 - g z|_{z = h({\bf{r}}_{H},t)} \mathrm{.}
\label{eq:Be2}
\end{equation}

 The other conditions to be satisfied by $\Phi$ are mass
conservation, which becomes Laplace's equation:
 \begin{equation}
\nabla^2 \Phi = 0
\mathrm{,}
\label{eq:Be3}
\end{equation}
and the Neumann  (no-flow) condition at the bottom (assumed for
simplicity at $z =0$),
\begin{equation}
 \frac{\partial \Phi}{\partial z}  = 0|_{z = 0}
\mathrm{.}
\label{eq:Be4}
\end{equation}
 Given the value of $\Phi$ on the free surface and the Neumann
condition at $z = 0$, one can solve Laplace's equation, once it is
transformed into a linear BIE. Concretely this
means that the normal derivative of $\Phi$ on the free
 surface can be found by solving an integral equation on the boundary, the remaining components
  of the gradient of $\Phi$, there, being known from the data. This allows
  us to compute the left-hand
  side of equation (\ref{eq:Be2}), from its right-hand side, and thus to advance the value of $\Phi$ on
  the free
  surface, in time.
  The evolution of the  free surface elevation is given by the
  kinematic condition that it be a material surface, thus
  advected by the local fluid velocity ($w$ is the vertical fluid velocity):
 \begin{equation}
\frac{\partial h}{\partial t} + \frac{\partial \Phi}{\partial x}
\frac{\partial h}{\partial x} + \frac{\partial \Phi}{\partial y}
\frac{\partial h}{\partial y}  = w|_{z = h({\bf{r}}_{H},t)}
\mathrm{.} \label{eq:Be5}
\end{equation}
Equations (\ref{eq:Be2}) and (\ref{eq:Be5}),  assuming that the
velocity components are computed via the solution of Laplace's
equation (\ref{eq:Be3}) (by a BIE), define a closed problem for the coupled evolution of
 the velocity potential and  the free surface. All of this, however,
 assumes that the free surface remains simply
 connected.
  The explicit form of the BIE problem is as follows. It uses the 3D free space
 Green's function defined as \cite{Brebbia},
\begin{equation}
G({\bf{x}}, {\bf{x}}_1) = \frac{1}{4\pi r} \mathrm{,} \label{eq:Be6p}
\end{equation}
with the normal derivative
\begin{equation}
\frac{\partial G({\bf{x}}, {\bf{x}}_1)}{\partial n} = -
\frac{1}{4\pi } \frac{{\bf{r}}\cdot{\bf{n}}}{r^3} \mathrm{,}
\label{eq:Be7}
\end{equation}
and $r$ being the distance from point ${\bf{x}} = (x, y, z)$ to the
reference point ${\bf{x}}_1 = (x_1, y_1, z_1 )$, both being on the
boundary, and $\bf{n}$ representing the outward normal unit vector to
the boundary at point ${\bf{x}}$. Green's second identity transforms
Laplace's equation into the BIE,
\begin{equation}
\alpha ({\bf{x}}_1) \Phi({\bf{x}}_1) = \int_{\Gamma({\bf{x}})}\left[
G({\bf{x}}, {\bf{x}}_1) \frac{\partial \Phi({\bf{x}})}{\partial n} -
\Phi({\bf{x}}) \frac{\partial G({\bf{x}}, {\bf{x}}_1)}{\partial
n}\right] {\mathrm{d}}\Gamma \mathrm{,} \label{eq:Be8}
\end{equation}
in which $\alpha ({\bf{x}}_1) = \frac{\theta_1}{4\pi}$, with
$\theta_1$ the exterior solid angle
 made by the boundary at point ${\bf{x}}_1$ (i.e.
$2\pi$ for a smooth boundary). Moreover $ {\mathrm{d}}\Gamma$ is the
element of area on the surface at position ${\bf{x}}$. The accurate
numerical solution of  BIE (\ref{eq:Be8}), with the bottom boundary
condition (\ref{eq:Be4}), together with the time integration of
Equations (\ref{eq:Be2}) and (\ref{eq:Be5}), allows simulating 3D
overturning waves such as depicted in Figure 2a \cite{GGD,GG}, up to
the time the plunging jet is about to impact the free surface
underneath it.
\subsubsection{ Generalisation of the Integral equation method}

 In the case of a plunging jet impacting the underlying surface,
the domain becomes multi-connected and the above BIE  cannot be used as such
 anymore. We discuss a way to extend this method to such cases.
 According to the Kelvin-Helmholtz theorem, for an inviscid fluid, no vorticity can be created outside of the
 domain boundary. In the present case, this applies to the boundary
  between the free falling overturning jet and the fluid underneath.
  Thus, assuming the jet does not immediately mixes with the impacted
 fluid (and this is in fact supported by some laboratory experiments for early
 times after impact) a vortex sheet
   can be specified on the surface separating the two fluids \cite{vortexsheet}, to
   simulate the tangential velocity discontinuity that appears,  such
 that the plunging jet may continue to exist as a region of potential flow
beyond the time of intersection.
    Thus the jet has two parts, an upper one in contact with  ambient air (where $p=0$) , and a lower part
    inside the underlying fluid, which is separated from the
    surrounding fluid by high pressure air (see Figure 8).
Let us now derive a closed set of dynamical equations for quantities
defined on surfaces, as in the previous case prior to intersection.
For that purpose, we introduce three boundaries, labeled $B_1$,
$B_2$ and $B_3$, respectively (see Figure 8). $B_1$ is the bottom at
$z =0$, $B_2$ is the free surface between the air and the fluid,
where the pressure is zero, and $B_3$ is the inner surface
separating the fluid coming from the jet from the underlying fluid.

\begin{figure}[ht]
\centerline{\includegraphics[scale=1]{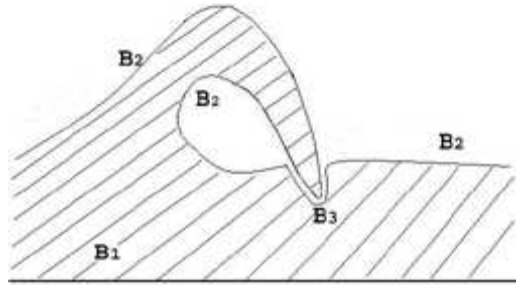}}
 \caption{ Scheme of the plunging jet}.
\end{figure}

 Because of the existence on $B_3$  of the vortex sheet, i.e., concentrated
vorticity, one cannot take as dynamical function the value of the
 velocity potential there, since it jumps across this interface.
 Therefore we shall take $\frac{\partial \Phi({\bf{x}})}{\partial
 n}$
 instead of $\Phi({\bf{x}})$, as dynamical variable on $B_{3}$ where
 the normal velocity and the normal acceleration have both to be
 continuous, to make the two sides of the boundary moving together.
 The BIE for $\Phi$ can be solved from the knowledge of $\frac{\partial \Phi}{\partial
 n}$ (Neumann condition) on $B_{1}$, $B_{2}$, and $\Phi$ on $B_{3}$
 (Dirichlet condition). This yields the set of boundary conditions
 for solving the Laplace equation (\ref{eq:Be3}) inside the fluid.
 To advance in time and derive the dynamical equation for $\frac{\partial \Phi}{\partial
 n}$ on $B_{3}$, let us  define $ q = \frac{p}{\rho} + \frac{1}{2}\, \mid\nabla \Phi\mid^2 $ that we
 shall call the dynamical pressure. By taking  the Laplacian of the Bernoulli equation (\ref{eq:Be1}),
 and because $\Phi$
  and $z$ are harmonic functions, one finds:
\begin{equation}
\Delta q  = 0 \mathrm{.} \label{eq:Be6}
\end{equation}

We are now ready to solve equation (\ref{eq:Be6}), using only two
 boundary conditions for $q$, $q = \frac{1}{2}\, \mid\nabla \Phi\mid^2$ on the free surface
 $B_{2}$ where $ p=0$, plus
the condition $\frac{\partial \left( q - \frac{1}{2}\, \mid\nabla
\Phi\mid^2 \right) }{\partial z} +  g = 0$ on
 the bottom $B_{1}$. Indeed the portion $B_{3}$ does not enter
  in the contour of the BIE for $q$, because both $q$ and $\frac{\partial q}{\partial
 n}$ are continuous on $B_{3}$ for the following reasons.

Defining the velocity as $\bf{\overrightarrow{u}} =
-\bf{\overrightarrow{\nabla}} \Phi$, the Euler fluid equation can be
written as,
\begin{equation}
 \frac{\partial {\bf{\overrightarrow{u}}}}{\partial t} + \bf{\overrightarrow{\nabla}} q + g
{\bf{e}}_z = 0 \label{eq:uq} \mathrm{,}
\end{equation}
 where ${\bf{e}}_z$ is the
unit vector in the vertical direction. The continuity of  $q$ across
$B_{3}$ follows from equation (\ref{eq:uq}) since otherwise an
infinite acceleration would take place. Moreover the continuity of
$\frac{\partial q}{\partial
 n}$ is ensured because of the continuity of the normal acceleration
across $B_{3}$. Therefore the Laplace equation (\ref{eq:Be6}) for
$q$ can be solved by using the boundary conditions on $B_{1}$ and
$B_{2}$ only.

   This means that, like
   the problem of evolution of the free surface in the simply connected case, the post-collision dynamical
   equations (for $\Phi$ and  $q$) can be reduced to linear integral
    equations posed on the boundaries ($B_{1,2,3}$ for $\Phi$, and $B_{1,2}$ for $q$).

Because of the merging of the plunging jet with the fluid
underneath, vorticity is generated in a
 flow that is otherwise potential. This vorticity is roughly horizontal in the span-wise direction
 of the waves. If those waves are more or less parallel to each other
(i.e., long crested) , this could be a coherent source
  of vorticity near the surface that would become on a large scale a shear layer driving the fluid
  opposite to the direction of the wave propagation.  Another difficulty in this problem is the possible
   occurrence of regions of negative pressure where nucleation of bubbles should take place (actually of
   pressure less than the value of saturating pressure for the given temperature condition).  We
   set
   aside this possibility, although it is obvious from observations that air bubbles are generated in
   white caps and even more so by plunging breakers.

\subsubsection{Hertz-fluid contact problem}
 A possible explanation for this generation of many
bubbles lies in a simple scaling law for the Hertz-fluid contact
problem. Let us assume that the plunging jet has a locally parabolic
tip hitting the fluid underneath, for simplicity,  assumed to be
flat. Therefore we have a typical length scale, the radius of
curvature of the parabolic tip $R$, and
 a velocity scale, the velocity $w$ of the falling jet when it hits the flat surface.
 Because  Bernoulli equation has no intrinsic scale, every parameter, like for instance
 the typical value of the velocity in the collision region (defined a bit more accurately below),
 is like ${\bf{u}}(x,t ) = w  {\bf{u}}_{sc}(x/R, wt/R)$, where the vector field ${\bf{u}}_{sc}(x/R, wt/R)$
 is dimensionless, universal, and defined by numerical functions only, $ t=0$ being
 defined as the time of the first
  contact. One may try to find the behavior of the various quantities involved just after the collision
  by borrowing some of the ideas of the famous Hertz contact problem \cite{Hertz} between solids.
  As in Hertz contact,
  one assumes that the perturbation comes from the flattening of the
  tip, wherever it would be inside the
  fluid underneath if this were possible. The tip would have penetrated by a length $wt$ inside this fluid,
   and so spread over a length of order $\lambda \sim (R wt)^{1/2}$ along this surface. Because this brings
    a perturbation of order 1 to the boundary of the fluid inside the jet and because Laplace's equation
     has
     no typical length scale (this argument is directly borrowed from
     Hertz's contact problem, where Hertz used this property of the equations of  elasticity),
     the region perturbed inside the jet
      is a volume of size $\lambda$ in all directions. The vertical momentum of the falling jet in this
       perturbed region is therefore of order
\begin{equation}
P_{z} \sim \rho w \lambda^3
 \mathrm{.} \label{eq:H1}
\end{equation}

       The
    change with time of this momentum $\frac{dP_{z}}{dt}= p\lambda^{2}$ is due to the pressure $p$
    near the boundary between the
    falling jet and the fluid underneath over an area $\lambda^2$. This yields an order of magnitude
     for the pressure in the collision domain
\begin{equation}
p \sim \rho w^2 \left(\frac{R}{wt}\right)^{1/2}
 \mathrm{,} \label{eq:H2}
\end{equation}

      when using the approximation $ \frac{dP_{z}}{dt}\sim
      \frac{P_{z}}{t}$. The relation (\ref{eq:H2})
      shows that this pressure, felt across the boundary $B_{3}$,
       diverges at the instant of contact where it is regularized probably by capillary
       phenomena and/or compressibility effects in the fluid (in the compressible case, at short time, the divergence of pressure is replaced
       by a very large peak of pressure of order $\rho w c_{s}$, with $c_{s}$ the speed of sound in the fluid). This result could explain
       in part the formation of
     bubbles and foam in the real world.

       \section{Final remarks}
       \label{sec:conclusion}

        We presented a rather simplistic, but physically
       meaningful, approach to wave-breaking,  yielding scaling laws
       that were shown to be in good agreement
       with numerical results for wave overturning in 3D, based on
       a higher-order
       solution of full dynamic equations. This work is an example of
       complementarity between advanced
       numerical and analytical methods, each providing validation
      and physical meaning to the other. Analytical methods also
      make it possible to generalize nonlinear properties observed in
      numerical results, to a whole class of equations and
      problems.

      In particular, and on a more general point of view, it seems to us that many important
       concepts and ideas pertaining to nonlinear science remain to be
       exploited, in many areas where they can
        guide the intuition or better yield specific predictions that can
        then be compared to
        numerical results. In the specific application of
        nonlinear fluid mechanics presented here, we have
       encountered various more generic subproblems, all with a definitely very strong nonlinear flavor,
       like the singular behavior of solutions of partial differential
       equations, or what we
        referred to as the Hertz-fluid contact problem. In
        closing, this
        indicates  that free boundary problems
         remain a rich and perhaps inexhaustible source of inspiration in the field of nonlinear science.

\begin{acknowledgments}
 Paul Clavin, Christophe Josserand, Laurent Di Menza and Fr\'ed\'eric Dias are
gratefully acknowledged for stimulating discussions.
 Philippe Guyenne acknowledges support from
the University of Delaware Research Foundation and the US National
Science Foundation, under grant DMS-0625931. St\'ephan Grilli
acknowledges support from the U.S. ONR Coastal Geosciences Division
(code 321CG) Grant No. N000140510068.

\end{acknowledgments}

{}

      \ifx\mainismaster\UnDef%
      \end{document}
:47.10g; 47.35jk; 05.45a